# Step-by-Step Guide to Conducting Meta-Analysis of Dichotomous Outcomes Using RevMan in Dental Research

Hoi-Jeong Lim
*Department of Statistics,
Department of Orthodontics, Chonnam National University School of
Dentistry, Dental Science Research Institute
Chonnam National University
Gwangju, South Korea*

Su-Hyeon Park
*Department of Statistics,
Department of Orthodontics, Chonnam National University School of
Dentistry, Dental Science Research Institute
Chonnam National University
Gwangju, South Korea*

*Abstract*— Meta-analysis is a statistical method that combines the results of individual studies on the same topic. This method is becoming popular, due to providing the combined result that individual studies cannot provide and giving a more precise result. Despite meta-analysis having such significance, there are few Korean guides for the use of the Review Manager (RevMan) software. This study will provide a step-by-step guide, using orthodontic mini-screw as a dental example, to help researcher carry out meta- analysis more easily and accurately.

*Keywords*— Meta-analysis; RevMan; dental research

## I. Introduction

Meta-analysis refers to a statistical method used to objectively summarize the results of various prior studies conducted on a specific research topic into a common effect size. The advantages of meta-analysis include increased statistical power, the ability to draw more systematic and accurate conclusions than individual studies, and the capacity to identify the causes of conflicting research results. However, its disadvantages may include issues arising from differences in the quality of the studies included in the analysis, and the potential lack of representativeness since only published studies are selected[1].

In a randomized study[2], most systematic reviews and meta-analyses of dental research papers have addressed general introductions and procedures related to these topics, interpreting the results found in published meta-analysis papers from a statistical perspective. Additionally, various other papers[1, 3] also cover the overall content of meta-analysis. Therefore, this study aims to investigate how meta-analysis was actually conducted in dental field meta-analysis papers using examples, employing the open-source software ReviewManager (RevMan). The purpose of this study is to demonstrate step-by-step the procedures necessary for conducting a meta-analysis, such as creating a PRISMA diagram, tables regarding the characteristics of included studies, risk of bias summary, forest plot, and funnel plot, in a way that even non-statisticians can easily follow.

## II. Information on the Dataset Used Here

To learn how to use RevMan software step by step for meta-analysis, data from a published meta-analysis paper in the field of dentistry (Hong (2016))[4] was utilized. Recently, there have been active studies on mini-screws[5-24], and meta-analysis studies combining these individual research findings are being published. One such study by Hong[4] employed a meta-analysis method that combined various research results regarding whether the placement location of mini-screws (maxilla vs. mandible) affects the success of mini-screw placement. This study aimed to demonstrate how to calculate a common effect size using these individual research results, create a forest plot and funnel plot, and draw a PRISMA diagram, Characteristics of included studies table, and Risk of bias summary. The goal was to explore how the various results obtained from the study by Hong were carried out step by step using RevMan.

## III. Step-by-step method for conducting meta-analysis using RevMan

The step-by-step method for conducting a meta-analysis using RevMan is as follows: First, install RevMan 5. Second, set a title regarding what type of meta-analysis will be performed on the systematically reviewed papers. Third, create a PRISMA diagram that explains how the papers were selected. Fourth, add the papers to be analyzed in RevMan. Fifth, create a table listing the general characteristics of the individual studies and draw a Risk of Bias summary to evaluate the quality of the papers. Lastly, create a Forest plot to show the overall effect size through the meta-analysis and a Funnel plot to indicate publication bias[Fig. 1].

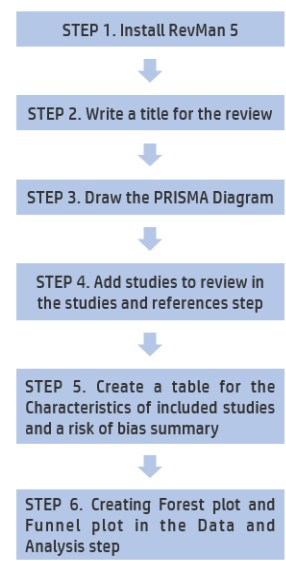

Fig. 1. A step-by-step guide to Meta analysis

*A. Step 1: Installing RevMan 5*

To install the RevMan5 software from the Cochrane Community[25], click the link below to see the screen in "Fig. 2" and download the file that is suitable for your computer system. Once the installation is complete, the following icon will be created on your desktop. Double-clicking this icon will open the window shown in 2-1 of "Figure 3." http://community.cochrane.org/tools/review-production-tools/revman-5/revman-5-download [Fig. 2].

*B. Step 2: Writing a Title for a Review*

To create a title for a meta-analysis comparing the stability of miniscrews in the maxilla and mandible, titled "Maxilla vs. Mandible for Miniscrew Stability," the following process is followed to generate the title in the Content window [Fig. 3].

*1) In the "Welcome to Review Manager 5.3" window, click the Close button corresponding to ① and then click the New icon corresponding to ②*

*2) When the New Review Wizard window opens, click the Next button.*

*3) When a window opens asking you to select the type of Review you want to create, select the default option, Intervention review, and click the Next button.*

*4) For a study comparing the success rates of mini-screw placement in the maxilla and mandible, enter the title as shown in Figure 3 of 2-4, and click the Next button.*

*5) When a window opens asking which Stage to start with, select Full review and click the Finish button.*

*6) Then, a window will open divided into three sections: the Outline window on the left, the Content window in the middle, and the Guidance window on the right, as shown in 2-6.*

*C. Step 3: PRISMA Diagram*

PRISMA diagram is a flowchart that describes how systematically reviewed papers were collected through various stages. It includes the number of identified papers, the number of included or excluded papers, and the reasons for exclusion[Fig. 4].

*1) Right-click on the Outline pane and select Add Figure.*

Fig. 2. How to install RevMan 5 for step 1

*2) This will open the New Figure Wizard window. Select Study flow diagram (PRISMA template) under Figure Type and click the Next button. A window will open for entering a caption; enter an appropriate caption and then click the Finish button.*

*3) A new PRISMA template will be created in the Content pane. To edit the content inside the box, double-click the box, which will open the Edit flowchart box. After making the necessary changes, click the OK button.*

*4) Once the modifications are complete, a PRISMA diagram like the one in step 3-4 will be generated.*

*D. Step 4: Adding studies and references*

There are two ways to add papers for conducting a meta-analysis. The first method is to manually input each paper's author, title, journal name, publication year, volume, pages, etc., as shown in "Fig. 5". The second method is to search for papers on PubMed, save them as a txt file, and then import them into RevMan for use, as shown in "Fig. 6"[Fig. 5, 6].

*1) To manually input each paper's author, title, journal name, publication year, volume, pages, etc.*

   *a) In the Outline pane, double-click on Studies and references > References to studies, then click on Included studies under References to studies. This will display Included studies in the Content pane. Click the Add Study button below it.*

   *b) When the New Study Wizard window opens, enter the study relevant to this topic, Antoszewska 2009[15], and click the Next button. A window will appear for selecting the type of Data Source; choose Published and unpublished data and then click the Next button. In the next window, confirm the year 2009, then click the Next button. When the Add Identifier window opens, click the Next button. If you have more studies to add, select Add another study in the same section, then click Continue. If there are no more studies to add, select Nothing and then click Finish. You can add the remaining studies using the method in 4-1-2.*

   *c) In the Content pane, double-click on [Empty] under Antoszewska 2009[15] to open a window where you can enter details about the study.*

   *d) Study: After entering the three sub-items for Antoszewska 200915, which are Authors, English Title, Journal/Book/Source, Date of Publication, Volume, and Pages, click the Add Identifier button at the bottom right.*

   *e) Select an appropriate item from the drop-down list under Type in the bottom left corner for the type of Identifier.*

   *f) By clicking on Text of Review in the top left corner of the Content window, you can check the details entered for each study under Included studies.*

*2) To search for papers on PubMed, save them as a txt file, and then import them into RevMan for use.*

   *a) Go to the PubMed website at https://www.ncbi.nlm.nih.gov/pubmed/, log in, and find one of the papers identified through the PRISMA diagram by clicking on the title of the paper. When you select "Create collection..." from the drop-down list under "Save items" in the middle right of the screen, the screen for 2-b will appear.*

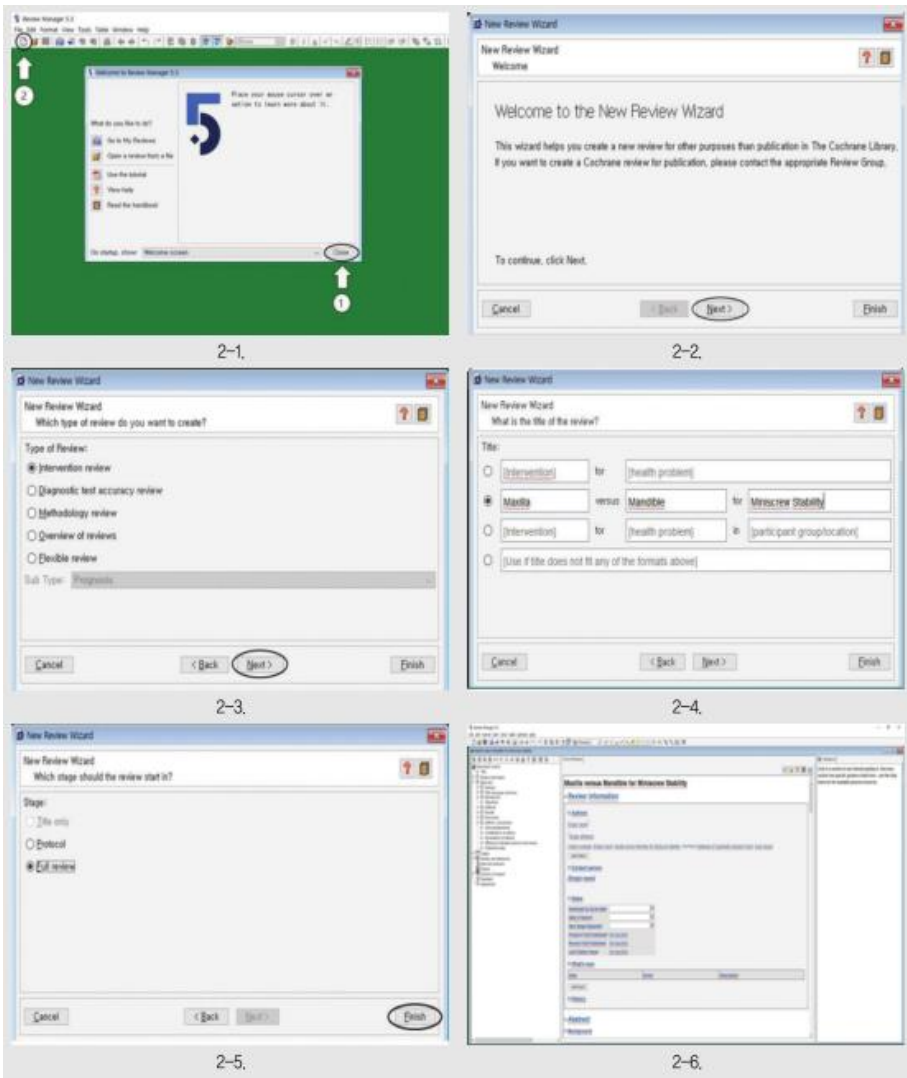

Fig. 3. How to write Title of review for step 2

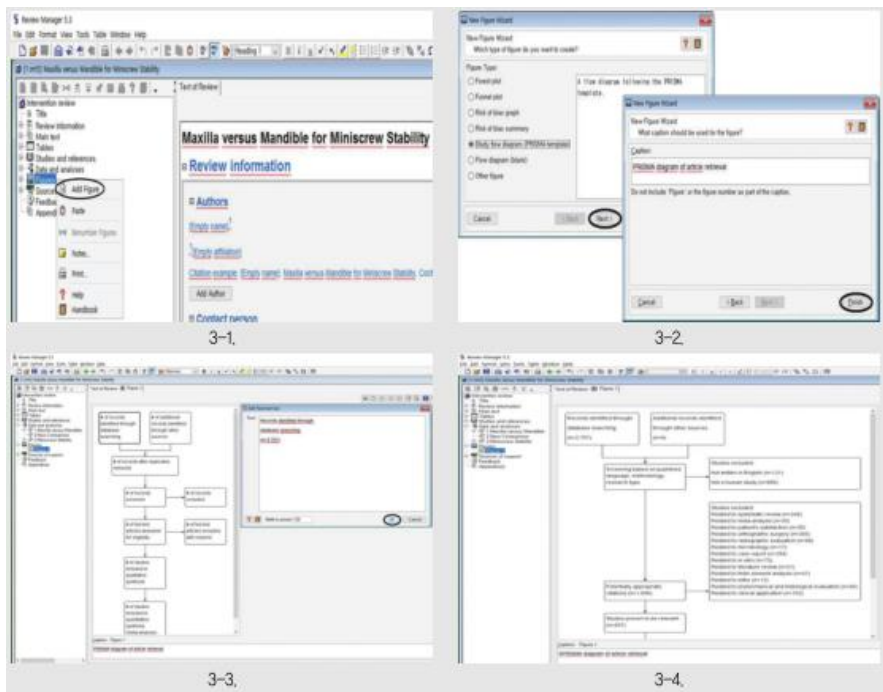

Fig. 4. PRISMA diagram for step 3

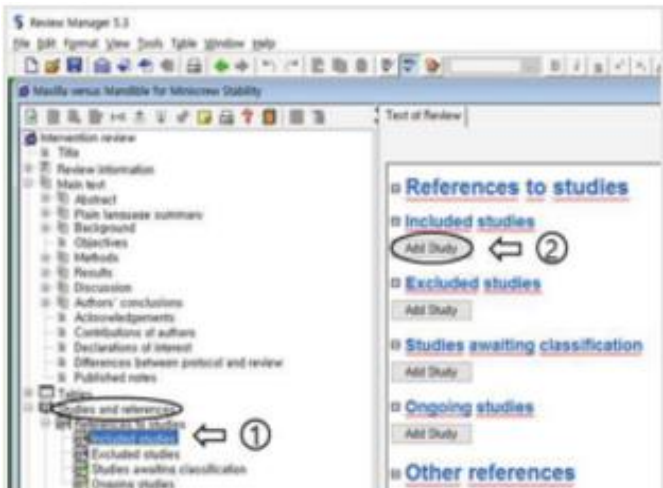
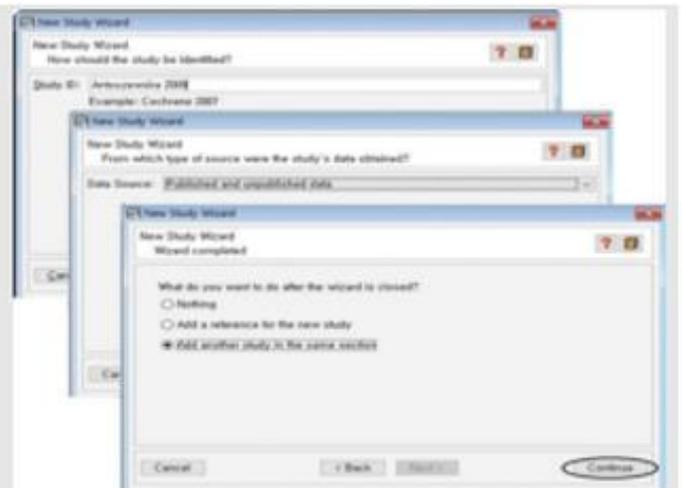
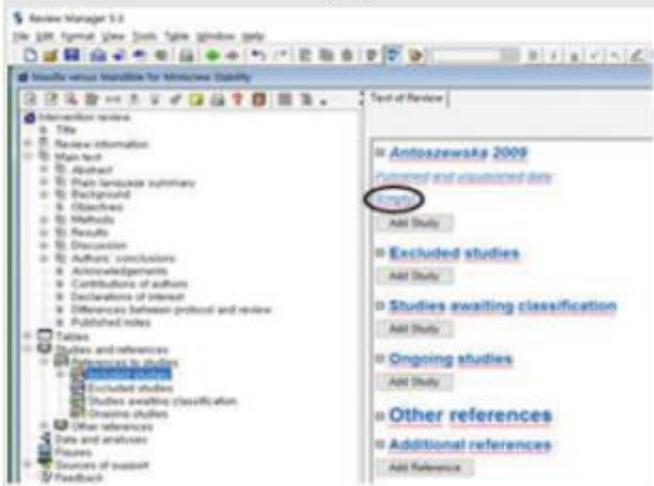
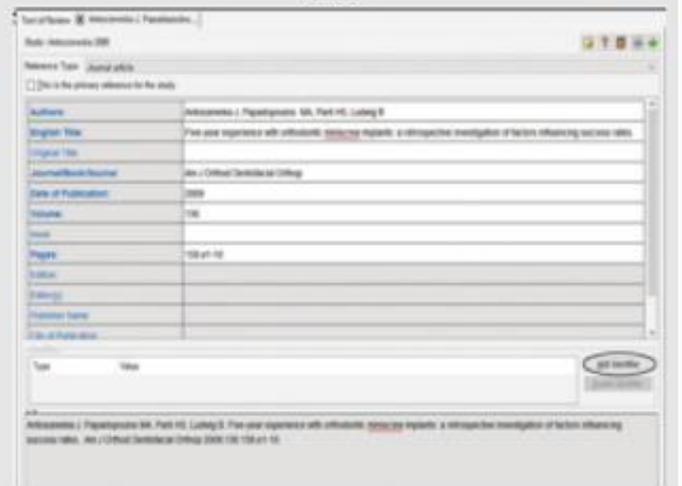
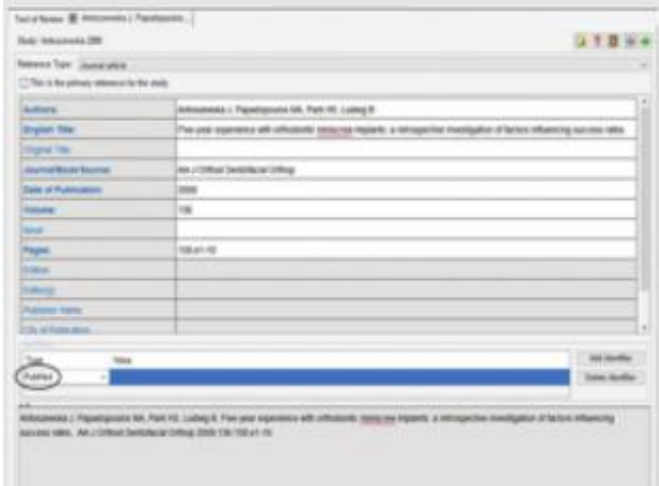
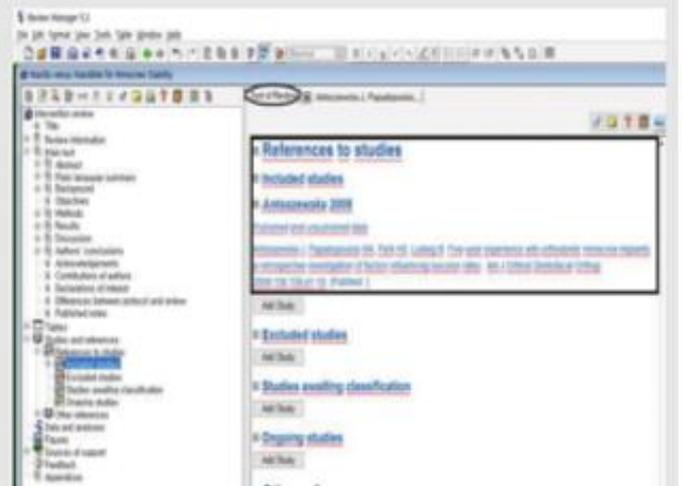

Fig. 5. PRISMA diagram for step 3

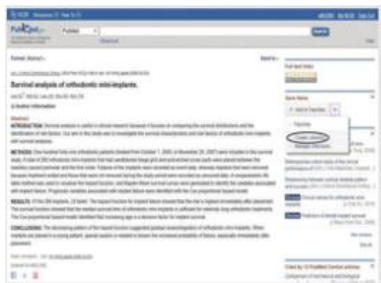
2-a

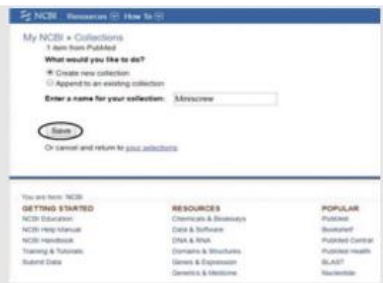
2-b

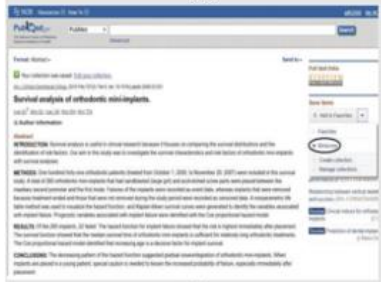
2-c

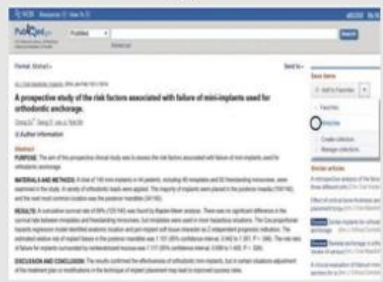
2-d

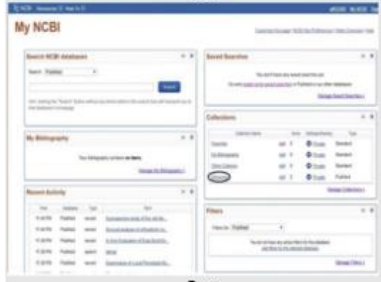
2-e

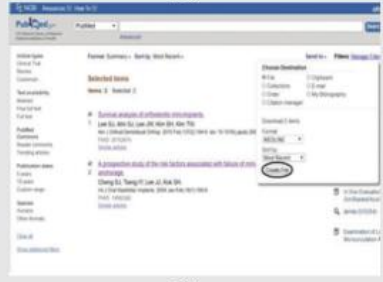
2-f

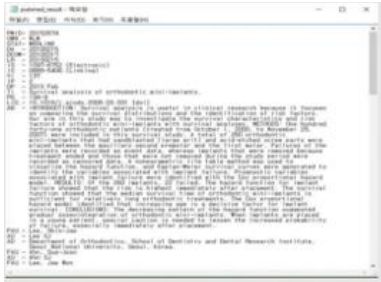
2-g

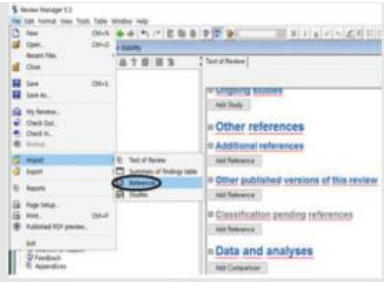
2-h

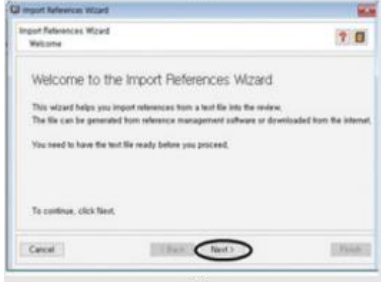
2-i

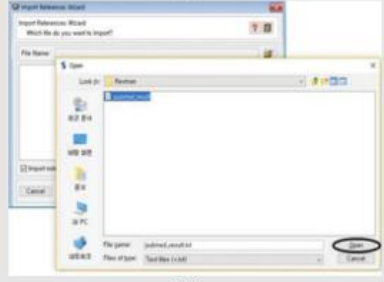
2-j

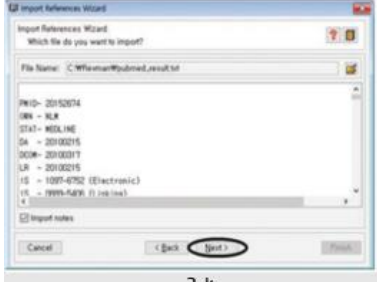
2-k

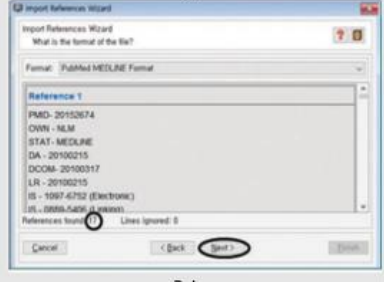
2-l

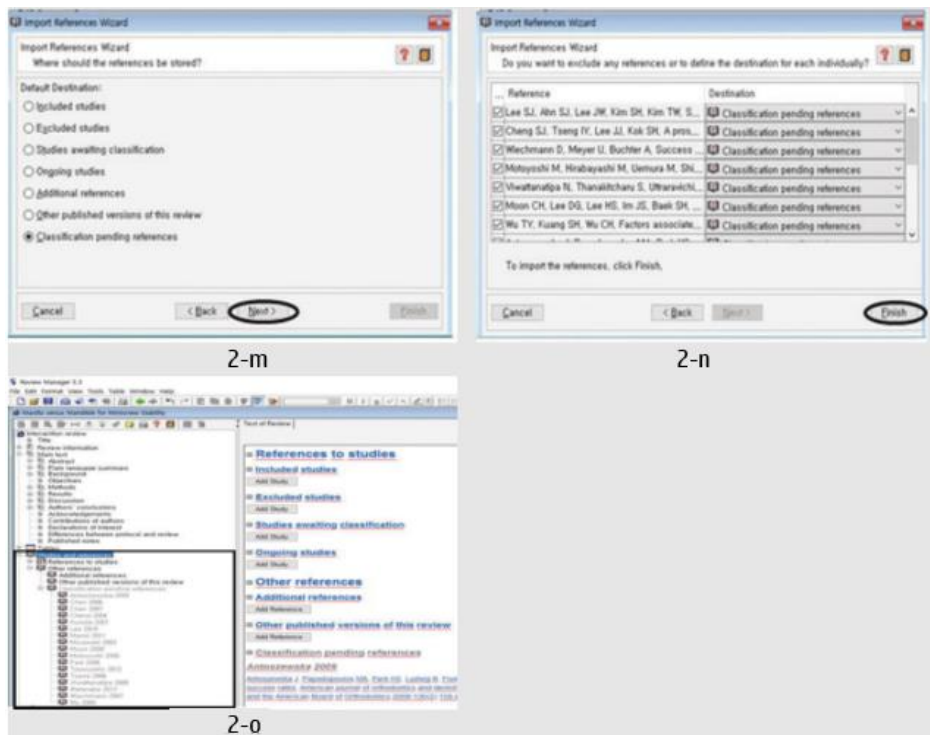

Fig. 6. Importing references from a text file for step 4

*b)* When asked what you want to do, click "Create new collection" and enter the name "Miniscrew" in response to the question asking for a name for your collection, then click "Save."

*c)* You will then return to the screen from 2-a, and although it is lengthy, you can confirm that a directory named "Miniscrew" has been created under "Save items," and that the paper found in 2-a has been added to that directory.

*d)* Find another paper, click on the title, and using the same method, click on the directory named "Miniscrew" under "Save items" to add the paper to that directory. In this way, you can save a total of 17 papers needed for the meta-analysis.

*e)* Click on "My NCBI" at the top right of the screen, and the screen for 2-e will appear. Click on the directory "Miniscrew" under Collections to proceed to the screen for 2-f.

*f)* After checking the small box next to each paper title, click the arrow next to "Send to," check "File," check "MEDLINE" under Format, and then click the "Create File" button.

*g)* When you open the generated text file, you will find the necessary information such as titles, abstracts, and authors of the previously added papers written in MEDLINE format, and save that txt file. The txt file is saved at the following address:

http://dent.jnu.ac.kr/user/indexSub.action?codyMenuSeq=6806&siteId=dent&menuUIType=top&dum=dum&boardId=334&page=1&command=view&boardSeq=570504&chkBoxSeq=&categoryDepth=&status=&moveUrl=dent.jnu.ac.kr

*h)* To import that txt file into RevMan, click on File > Import > References in the RevMan menu bar, and the Import References Wizard window for 2-i will appear.

*i)* In the Import References Wizard window, click the "Next" button.

*j)* A window will open asking what type of file you want to import, and at the same time, a window will appear allowing you to select the location of the saved file. Select the corresponding file and click the "Open" button.

*k)* If the path and content of the txt file are visible, click the "Next" button.

*l)* After confirming that the number in "References found" matches the number of papers you want to add, click the "Next" button.

*m)* When a window appears asking where the papers you want to add should be stored in RevMan, confirm the default option "Classification pending references" and click the "Next" button.

*n)* After confirming that the 17 papers are stored in the "Classification pending references" location, click the "Finish" button.

*o)* In the Outline window, you can confirm that the papers have been added under Studies and references > Other references > Classification pending references.

*E. Step 5: Tables*

Systematic reviews and meta-analyses typically include two tables. One table lists general characteristics extracted from the research papers, such as publication date, age and gender of the study subjects, study design, and sample size. The other table evaluates the quality of the included individual studies based on factors such as the type of study, presence of a control group, appropriateness of outcome variable selection, adequacy of sample size, measurement error assessment, and appropriateness of statistical methods. We aim to create a table that extracts the general characteristics of the research papers using RevMan and to illustrate the Risk of Bias Summary that evaluates the quality of the studies[Fig. 7].

The method for creating the Characteristics of Included Studies table is as follows:

- In the Outline window, double-click on Tables > Characteristics of studies > Characteristics of included studies > Antoszewska 2009[15]. Then, refer to the Methods, Participants, Interventions, and Outcomes sections of the table under Antoszewska 2009[15] in the Content window and fill them in based on the abstracts and results of the papers in sections 5-2 and 5-3.

- In the abstract of the paper by Antoszewska[15], it is reported that the research method was a retrospective investigation, so we entered "Retrospective study" in RevMan. It was reported that 130 patients with 350 mini-screws were the subjects of the study, so we entered "350 MIs in 130 patients" in RevMan. The outcomes were the success rates of the mini-screws, which we entered as "Success rates of MIs" in RevMan

- It is reported in Table 1 that 173 mini-screws were placed in the maxilla and 177 mini-screws were placed in the mandible, so we entered "173 MIs in the maxilla; 177 MIs in the mandible" in the Interventions section of 5-1.

The method for creating the Risk of Bias Summary is as follows:

- We reconstructed Table 2 from the paper by Hong(2016)[4] to fit the Risk of Bias Summary, including six items such as the adequacy of case definition using the Newcastle-Ottawa Scale for the 12 individual studies[13-24]. We aim to create the Risk of Bias Summary in 5-11 using this table.

- By clicking the gear-shaped button on the right side of the Risk of Bias table in 5-1, the Characteristics of Included Studies Properties window in 5-5 opens. Since all seven items selected as default in RevMan do not match the items in the table from 5-4, we click each of the seven items and uncheck the Activated box to deactivate them.

- Click the Add button at the bottom left and enter "Is the case definition adequate?" in the Bias column, then click the OK button. A new item titled "Is the case definition adequate?" will appear in the Risk of Bias tables. Similarly, we add the remaining items.

- The order of the items can be adjusted using the Move Up and Move Down buttons.

- Select the Authors' judgement (Low/Unclear/High) evaluated in the paper by Antoszewska(2009)[15]. In the case of Unclear, enter the basis for judgement in the Support for judgement field (for example, "Not reported"). If not entered, it will remain blank in the Risk of Bias Summary in below. The remaining 11 studies by Chen(2006)[14-24] will also have their Authors' judgement selected in the same manner.

- Right-click on Figures in the Outline window and click Add Figure.

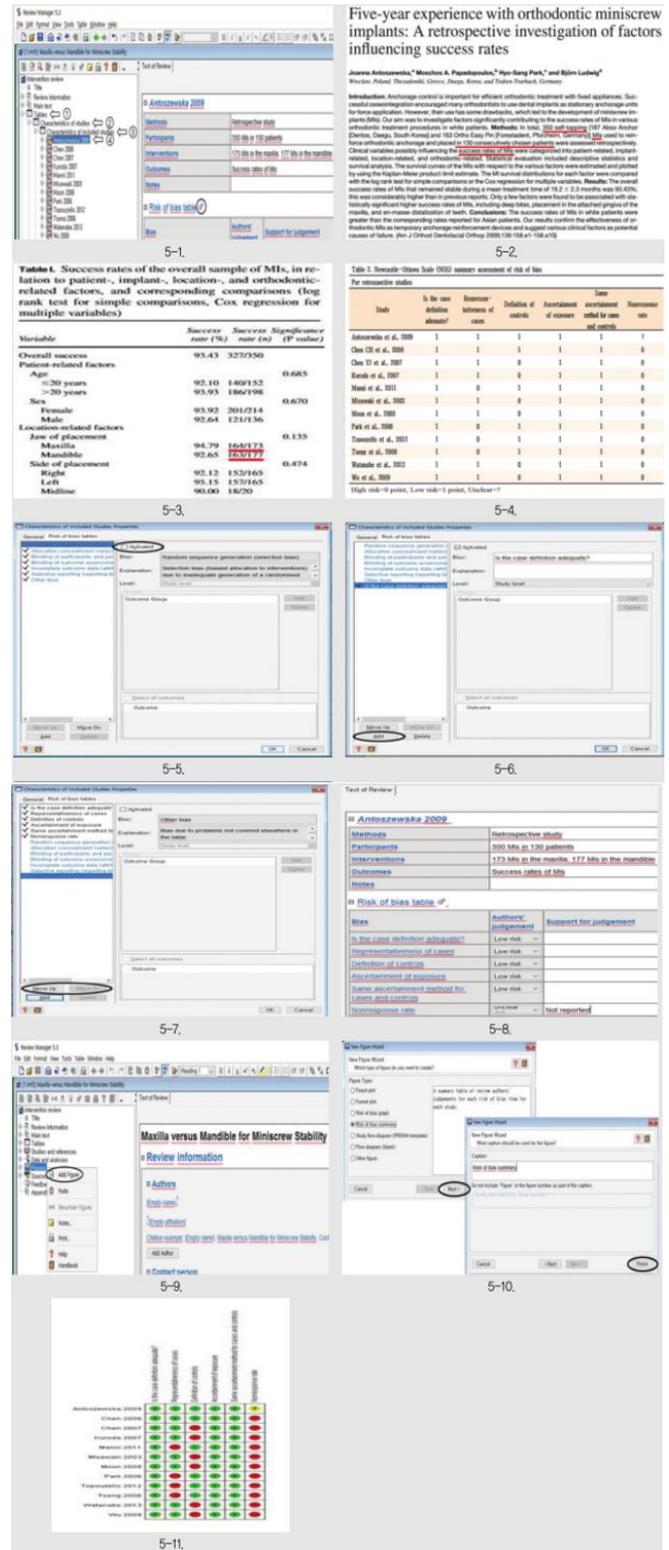

Fig. 7. Risk of bias summary for step 5

- Then the New Figure Wizard window will open, and by selecting "Risk of bias summary" in Figure Type and clicking the Next button, a window will open for entering a caption. After entering an appropriate caption, click the Finish button.

- The Risk of bias summary will be drawn as shown in 5-11.

*1) Risk of bias:* In RevMan, the risk of bias inherent in each study was evaluated using seven criteria: random sequence generation, allocation concealment, blinding of participants and personnel, blinding of assessors, incomplete outcome data, selective reporting, and other threats to validity. In the study by Hong(2016)[4], the Newcastle-Ottawa scale was used to categorize non-randomized studies into three groups (selection of study groups, comparability of groups, and confirmation of exposure or outcomes in case-control or cohort studies) based on eight items to assess the risk of bias inherent in each study. In this study, "Table 2" of Hong(2016)[4] was reconstructed into a quality assessment table with six items, and a Risk of Bias Summary was created using RevMan.

In the Risk of Bias Summary of RevMan, the quality of the studies was assessed in terms of high risk, unclear, and low risk for each bias risk. Among the six items used to evaluate the risk of bias, low risk results were obtained from each study for three items: appropriateness of case definition, confirmation of exposure, and the same confirmation method for both cases and controls. The representativeness of cases was reported as low risk in eight studies, and the definition of controls was also assessed as low risk in six studies. The non-response rate was reported as high risk in all studies except for one, and the study by Antoszewska (2009)[15] was classified as unclear due to the non-response rate being unclear. (It should be noted that "Fig. 7", 5-4 is a table created arbitrarily.)

*F. Step 6: Forest plot and Funnel plot*

A forest plot is a diagram that displays the most important results of a meta-analysis, showing the comprehensive results along with the results of several individual studies addressing the same question. In a forest plot, the size of the squares representing each individual study is determined by the relative weight or sample size, and the horizontal lines next to the squares indicate the 95% confidence intervals. The overall effect size is represented in the shape of a diamond, where the center of the diamond indicates the combined overall effect size, and the horizontal diagonal lines of the diamond represent the 95% confidence interval of the overall effect size. If this confidence interval includes 0, it indicates that there is no significant difference, while if it does not include 0, it signifies that there is a significant difference.

The funnel plot, commonly used in systematic reviews and meta-analyses, is a diagram that can confirm the presence or absence of publication bias. When the white dots in the funnel plot are symmetrically arranged, it indicates that there is no publication bias[Fig. 8].

- In the Outline window, click on Data and analyses once, and in the Content window, click the Add Comparison button under Data and analyses.
- The New Comparison Wizard window will open; enter the title of the Comparison as Miniscrew Stability and click the Next button.
- After the New Comparison Wizard window closes, a prompt will appear asking what you would like to do next. Select Add an outcome under the new comparison to add an Outcome to the Comparison created in step 6-2, and click the Continue button. This will open the New Outcome Wizard window.
- For the Outcome's Data Type, if the Outcome is bonding time, select Continuous; however, if you are separating the maxilla and mandible as in the example of the Miniscrew used above, select Dichotomous and click the Next button.
- Enter Miniscrew Stability as the name of the Outcome and Maxilla and Mandible as the Group Labels, then click the Next button.
- Among the analysis methods, select Mantel-Haenszel for the Statistical Method, Fixed Effect for the Analysis Model, and Odds Ratio for the Effect Measure, then click the Next button. (Detailed explanations of the selection methods are provided below "Fig. 7")
- In the analysis details, select Totals and subtotals for Totals, and 95% for both Study Confidence Interval and Total Confidence Interval, then click the Next button.
- For the graph details, enter Higher Success (Mandible) for the Left Graph Label and Higher Success (Maxilla) for the Right Graph Label, then click the Next button.
- To add study data for the new outcome, select Add study data for the new outcome and click the Continue button.
- Since there were 11 studies that clearly reported the success rates of implants in the maxilla and mandible, hold down the Shift key and click to select these 11 studies, then click the Finish button.
- A workspace will open in the Content window to draw a Forest plot to calculate the Overall effect size of the meta-analysis comparing the implant locations of the maxilla and mandible with Miniscrew Stability as the outcome variable.
- Looking at Table 1 in the paper by Wu(2009)[14], among the 135 miniscrews implanted in the mandible, 118 were successful, and among the 268 miniscrews implanted in the maxilla, 243 were successful.
- Looking at Table III in the paper by Miyawaki(2003)[17], among the 61 miniscrews implanted in the mandible, 51 were successful, and among the 63 miniscrews implanted in the maxilla, 53 were successful.
- In the Events column of the table in the Content window, enter the number of successful miniscrews implanted in the maxilla or mandible for each study, and in the Total column, enter the total number of miniscrews implanted in the maxilla or mandible. The respective Effect sizes and Overall effect size will be automatically calculated, and the Forest plot will be drawn.

- Clicking the Funnel plot button in the upper right corner will open the Funnel plot window, displaying a Funnel plot that indicates publication bias.
- When selecting the effect size in step 6-6, the calculation formulas for each effect size are provided in step 6-16, and detailed explanations are given below "Fig. 8".

6-1.

6-2.

6-3.

6-4.

6-5.

6-6.

6-7.

6-8.

6-9.

6-10.

6-11.

6-12.

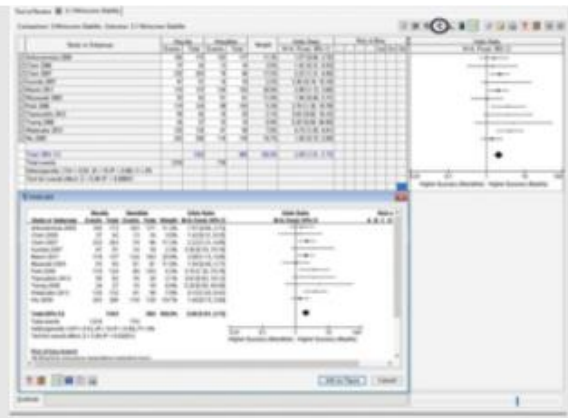
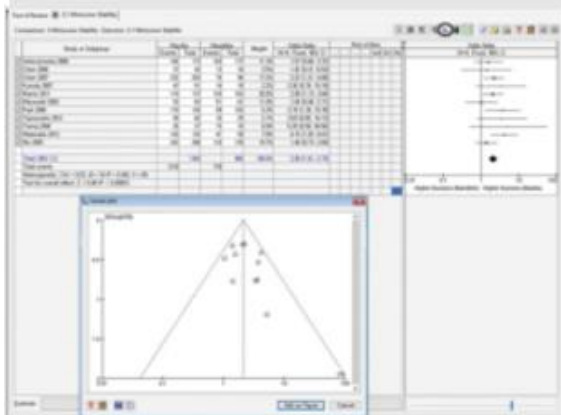
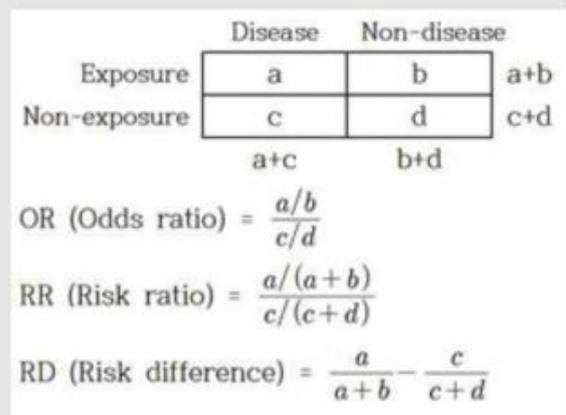

Fig. 8. Forest plot and Funnel plot for step 6

*1) Analysis models used in meta-analysis(6-6)*

The analysis models of meta-analysis include the fixed effect model and the random effects model[26]. While both the random effects model and the fixed effect model provide the same results when there is no heterogeneity among studies, when heterogeneity is present, using the random effects model results in a wider confidence interval for the average treatment effect and a more conservative statistical significance compared to the fixed effect model

*a) The Fixed Effect Model*

The fixed effects model assumes that the actual treatment effects assumed in each study are the same. However, the results appear differently because it is assumed that the results randomly occur around a common actual treatment effect. There are three representative statistical methods in the fixed effects model.

- Inverse variance method: The most commonly used method is to combine research results by using the inverse of the variance of the effect size as weights. This method is effective when the number of studies to be combined is small, but the sample sizes of each study are large.

- Mantel-Haenszel method: When the event rate is low or the study size is small, the estimate of the standard error of the effect estimate used in the inverse variance method increases, leading to a decrease in the accuracy of the effect estimate. However, the Mantel-Haenszel method is preferred over the inverse variance method because it uses different weights depending on the type of effect measure. This method is effective when the sample sizes of each study are small, but there are many studies to combine.

- Peto's method: Peto's method is an effective approach when the event rate is low or there are zero cells, but it has received a lot of criticism because it can produce biased odds ratios (OR) in studies where there is a significant difference in sample sizes between the case group and the control group.

*b) The Random Effect Model*

It is a model that assumes that the actual treatment effects differ because each study has different characteristics. Among the various treatment effects, the study weights are adjusted according to heterogeneity.

In the random effects model, two statistical models are provided: the Mantel-Haenszel method and the inverse variance method. The former estimates the amount of variance within a single study (between studies) by comparing the results of each study with the results of the Mantel-Haenszel fixed effects model meta-analysis, while the latter estimates the amount of variance across studies by comparing the results of each study with the results of the inverse variance fixed effects model meta-analysis. The difference between these two methods is very small (trivial).

*2) Interpretation of meta-analysis results (6-14, 15)*

Looking at the results of the meta-analysis, the homogeneity test results for checking whether the effect sizes of each study are homogeneous show that the null hypothesis is as follows, and the test statistic, degrees of freedom, p-value, and the heterogeneity index I2 were obtained as follows in 6-14.

$$H_o: OR_1 = OR_2 = \cdots = OR_{11}$$

$$Chi^2=9.53, df=10, p=0.48, I^2=0\%$$

The results of the homogeneity test showed that the p-value was greater than 0.1, and the I2 value was also 0%, proving that the effect sizes of each study were homogeneous. The combined effect size, represented by the Odds Ratio (OR), was 2.09, with a 95% confidence interval of [1.61, 2.73]. This means that the odds of success to failure in the maxilla (if the odds are 8, it indicates that for every 8 successes, there is 1 expected failure) were 2.09 times higher than the odds in the mandible. Additionally, the funnel plot is one of the methods to check for the presence of publication bias. Publication bias refers to the tendency of journal editors to prefer studies that show statistically significant differences, leading to a higher likelihood of such studies being published and included in meta-analyses, which can skew the results. A funnel plot is evaluated for the absence of publication bias if it has a symmetrical distribution regarding the combined effect size; in this case, it appears that there is little to no publication bias. If publication bias were present, it would be necessary to add studies that were likely not included through the trim-and-fill procedure to create symmetry and then recalculate the adjusted combined effect size.

*3) Effect size in binary outcome variables (6-16)*

*a) Odds ratio (OR)*

Odds refer to the ratio of the probability of contracting a disease (exposed group: a/(a+b), unexposed group: c/(c+d)) to the probability of not contracting the disease (exposed group: b/(a+b), unexposed group: d/(c+d)). The odds ratio refers to the odds of contracting the disease in the exposed group compared to the unexposed group.

*b) Risk ratio (RR)*

After observing a group exposed to risk and a group not exposed for several years, the ratio of the incidence rate in the exposed group compared to the incidence rate in the unexposed group is called the risk ratio or relative risk.

*c) Risk difference (RD)*

The difference in incidence rates between the exposed group and the unexposed group is called the risk difference.

*d) The difference between the three effect sizes*

RR is used only in cohort studies, while OR is more widely used in cohort studies, case-control studies, cross-sectional studies, and prevalence studies. RD is used in cross-sectional surveys or cohort studies, but it is not as widely used as RR or OR. Although there are differences among OR, RR, and RD as mentioned above, in RevMan, if you decide which of these three effect sizes to use, the values entered for Events or Total in 6-11 will be the same, and these three effect sizes will be calculated automatically.

## IV. CONCLUSION

Meta-analysis is a widely used technique not only in the field of dentistry but also in various other fields. As the number of studies on similar topics increases, it becomes difficult to identify trends, leading to a growing need for meta-analysis. RevMan is user-friendly for researchers who are new to meta-analysis and can be easily used for basic meta-analysis tasks. However, it does not support more complex meta-analysis functions such as meta-regression or network meta-analysis, so for advanced meta-analysis, software like Comprehensive Meta-Analysis is recommended. This study provides step-by-step guidelines for researchers who wish to conduct basic meta-analysis using the open-source software RevMan, and it is hoped that these guidelines will empower researchers to analyze independently and contribute, even in a small way, to deriving accurate research results.